\documentclass[letterpaper,journal]{IEEEtran}

\usepackage{amsmath}
\usepackage{amsfonts}
\usepackage{amssymb}
\usepackage{graphicx}
\usepackage{cite}
\usepackage{url}
\usepackage{booktabs}
\usepackage{tabularx}
\usepackage{array}
\usepackage{textcomp}
\usepackage{stfloats}
\usepackage{balance}
\usepackage[hidelinks]{hyperref}
\usepackage{orcidlink}

\newcolumntype{C}{>{\centering\arraybackslash}X}
\newcolumntype{L}{>{\raggedright\arraybackslash}X}

\begin{document}

\title{Probing Layer-Wise Robustness and Sensitivity of \\ Speech Enhancement Models Under Noise and Reverberation}
\author{Yair Amar, Amir Ivry\,\orcidlink{0000-0002-9043-7040},~\IEEEmembership{Senior Member,~IEEE} and Israel Cohen\,\orcidlink{0000-0002-2556-3972},~\IEEEmembership{Fellow,~IEEE}
\thanks{This research was supported by the Israel Science Foundation (Grant No.~1449/23) and the Pazy Research Foundation.}%
\thanks{The authors are with the Andrew and Erna Viterbi Faculty of Electrical and Computer Engineering, Technion -- Israel Institute of Technology, Haifa, Israel (e-mail: yairamr@gmail.com;  amirivry@gmail.com; icohen@ee.technion.ac.il).}
\thanks{This work has been submitted to the IEEE for possible publication. Copyright may be transferred without notice, after which this version may no longer be accessible.}%
}

\markboth{Preprint. Submitted to the IEEE for possible publication.}{Amar \MakeLowercase{\textit{et al.}}: Probing Layer-Wise Robustness and Sensitivity of Speech Enhancement Models}

\maketitle

\begin{abstract}
Speech enhancement (SE) models advance rapidly, yet how input degradation affects their internal representations remains underexplored. We introduce a probing framework to characterize how internal representations in SE models behave under controlled input degradation. We probe three SE models across controlled levels of signal-to-noise ratio (SNR) and reverberation, quantified by $C_{50}$, measuring layer-wise similarity to clean references with Centered Kernel Alignment (CKA) and summarizing each layer by a linear fit against degradation level: the intercept measures robustness, whereas the slope measures sensitivity. All three models are sharply non-uniform across depth, but they organize that non-uniformity differently. MUSE and MP-SENet grow more sensitive with depth, the sharpest transitions falling at MUSE's skip-connection junctions, where encoder information is reintegrated; Demucs inverts the trend. A randomly initialized model shows a near-flat profile, with slopes one to two orders of magnitude smaller, and the profile forms during fine-tuning, indicating that it is induced by the enhancement objective rather than a particular model design. Because CKA saturates at the clean reference, intercept and slope are partly coupled; we derive the identity relating them and report a \emph{saturation spread} statistic that indicates when their relationship is informative. Together, these results characterize where SE models are most sensitive to degradation. An exploratory analysis of whether residual variation tracks output-level quality, after controlling for SNR, shows that the speaker, rather than the utterance, must be treated as the sampling unit. Code and precomputed analysis artifacts for the main sweeps are publicly available.
\end{abstract}


\section{Introduction}

\IEEEPARstart{S}{peech} enhancement (SE) is a core component of many audio and communication systems, from hearing aids~\cite{green2022speech} and smart speakers to telecommunications front-ends such as acoustic echo cancellation~\cite{ivry2022objective} and automatic speech recognition pipelines \cite{loizou2013speech, vary2006digital}. Modern deep-learning-based SE models achieve strong output-level scores on metrics such as PESQ (Perceptual Evaluation of Speech Quality) \cite{pesq} and STOI (Short-Time Objective Intelligibility) \cite{taal2011stoi}. Yet, they are largely treated as black boxes: practitioners select and deploy models based on aggregate performance numbers without insight into how internal representations respond to varying input conditions. Understanding which layers preserve stable representations and which are most sensitive to degradation would complement output-level evaluation with a mechanistic account of how these models process degraded speech.

Probing methodology, introduced as linear classifier probes attached to the intermediate layers of image classification models \cite{alain2017probes}, has proven effective for revealing internal organization in other domains. In natural language processing, layer-wise probes applied to BERT \cite{bert} showed that successive transformer layers recapitulate a classical NLP pipeline, progressing from surface-level features to syntactic and semantic representations \cite{tenney2019bert, belinkov2019analysis, belinkov2022probing}. In computer vision, deconvolution-based visualization \cite{zeiler2014visualizing} and transfer experiments \cite{yosinski2014transferable} showed that early convolutional layers learn general features while deeper layers specialize in the training task, a pattern also captured quantitatively by representational similarity measures such as Singular Vector Canonical Correlation Analysis (SVCCA) \cite{raghu2017svcca} and Centered Kernel Alignment (CKA) \cite{kornblith2019similarity}.

This paradigm has been extended to audio. Layer-wise analyses of self-supervised speech foundation models, including wav2vec~2.0 \cite{baevski2020wav2vec}, HuBERT \cite{hsu2021hubert}, and WavLM \cite{chen_wavlm_2022}, have revealed a layer-wise hierarchy in which early layers capture acoustic features and middle-to-deep layers encode more linguistic information \cite{pasad2021layerwise, pasad2023comparative, pasad2024words, chung2021similarity}.
Layer choice has also been studied for enhancement directly: Tal et~al.~\cite{tal22_interspeech} conditioned a Demucs enhancement model on features drawn from individual HuBERT layers and found that the learned layer weights concentrate on the lower layers rather than on the mid-depth layers that score highest on phonetic probes, indicating that the layers most useful for enhancement are not those most useful for phonetic classification. However, these analyses target self-supervised models, whether evaluated on downstream classification tasks or used as a frozen feature source, leaving the internal representations of supervised SE models under-examined.

Santos and Falk~\cite{santos2018residual} compared residual and highway connections in SE models, characterizing residual connections as analogous to spectral subtraction and highway connections as combining spectral masking and subtraction, with visualizations of intermediate spectral-domain outputs. Mimilakis et~al.~\cite{mimilakis2019mapping} proposed a neural couplings algorithm to approximate and examine the learned filtering operators in denoising autoencoders for singing voice separation, characterizing the structure of inter-frequency interactions. Heitkaemper et~al.~\cite{heitkaemper2020demystifying} dissected Conv-TasNet by gradually replacing components of a frequency-domain baseline with time-domain counterparts to identify the functional contribution of each module to separation quality. Sivasankaran et~al.~\cite{sivasankaran2021explainable} applied DeepSHAP feature attribution to identify which time-frequency regions most influence the enhancement output. While each of these studies contributes to understanding SE model internals, none examines how layer-wise representations change systematically with input degradation severity.

We address this gap by probing the SE model MUSE \cite{lin24h_interspeech} under controlled degradation conditions, spanning additive noise with signal-to-noise ratio (SNR) from $-10$ to $30$\,dB and reverberation with clarity index $C_{50}$ \cite{ISO3382-1:2009} from $-5$ to $25$\,dB. We employ Centered Kernel Alignment (CKA)~\cite{kornblith2019similarity} to measure layer-wise representational similarity to clean references, and complement it with diffusion maps \cite{coifman2006diffusion, nadler2006diffusion}, a manifold learning technique that has been shown to reveal local intrinsic structure in audio representations~\cite{ivry2019voice, ivry2020evaluation,  ivry2026mapss}.

Our findings reveal a systematic \textit{robustness-sensitivity tradeoff} across depth: encoder layers maintain noise-invariant representations, whereas decoder layers are highly sensitive. Skip-connection boundaries mark the sharpest transitions. A similar trend emerges under reverberation and in MP-SENet \cite{mpsenet} and Demucs \cite{demucs}, two structurally distinct architectures, on both degradation axes, suggesting that the tradeoff is induced by the enhancement objective rather than by a particular model design. A randomly initialized MUSE, probed identically, produces a near-flat profile, and the profile forms during fine-tuning, which supports that reading directly. A diffusion-map view of the same representations shows the same depth asymmetry. We also examine the relationship between internal representations and output-level quality improvement, finding that the sampling unit for this question is the speaker rather than the utterance; that analysis is exploratory and describes the four speakers available to us.

The main contributions of this work are:
\begin{enumerate}
  \item A degradation-controlled CKA probing method for summarizing robustness and degradation sensitivity across SE model layers, together with a statistic that says when its two-parameter summary is informative.
  \item Evidence for a layer-wise, depth-organized robustness-sensitivity profile in MUSE under controlled degradation.
  \item Evidence that this pattern generalizes across additive noise, reverberation, and two additional SE architectures, MP-SENet and Demucs, and that it is induced by training rather than by architecture.
  \item An exploratory analysis of how layer-wise CKA relates to output-quality improvement, across three objective metrics.
\end{enumerate}
Code and precomputed analysis artifacts for the main sweeps are publicly available for reproduction.

The remainder of this paper is organized as follows. Section~\ref{sec:methods} describes the degradation model, the probed architectures and experimental setup, and the analysis tools. Section~\ref{sec:results} presents results under noise, the connection to perceptual quality, generalization to reverberation and across architectures, training-induced controls including dereverberation fine-tuning, and a geometric view from diffusion maps. Section~\ref{sec:discussion} discusses the findings in the context of classical spectral enhancement and their implications. Section~\ref{sec:conclusions} concludes the paper.

\section{Materials and Methods}
\label{sec:methods}

In this section, we describe the degradation axes, probed architectures, experimental setup, and analysis tools below.

\subsection{Degradation Model}
\label{sec:degradation_model}
We probe representations along two independent axes of input degradation: additive noise and reverberation, each swept from perceptually challenging to near-clean conditions.

\noindent For \textbf{additive noise}, the degraded signal at discrete-time sample $n$ is:
\begin{equation}
  y(n) = x(n) + v(n)
  \label{eq:additive_noise}
\end{equation}
where $x(n)$ is the clean utterance and $v(n)$ is a noise signal scaled to a target SNR. The SNR is swept in integer values from $-10$ to $30$\,dB. The corresponding clean utterances serve as the reference.

\noindent For \textbf{reverberation}, the degraded signal is:
\begin{equation}
  y(n) = \left(x * h\right)(n) = \sum_{k} h(k)\, x(n-k)
  \label{eq:reverb}
\end{equation}
where $*$ denotes convolution over time and $h(n)$ is a room impulse response (RIR). Reverberation severity is characterized by $C_{50}$, the logarithmic early-to-late energy ratio at the 50\,ms boundary~\cite{ISO3382-1:2009}, which correlates with perceptual quality and with speech recognition performance under reverberation more strongly than reverberation time and other standard room acoustic parameters~\cite{parada2016c50}.
For brevity, we drop the time-sample index $n$ in the remainder of the paper.

\subsection{Models and Activation Extraction}
\label{sec:probed_model}
\begin{figure}[t]
    \centering
    \includegraphics[width=0.62\columnwidth]{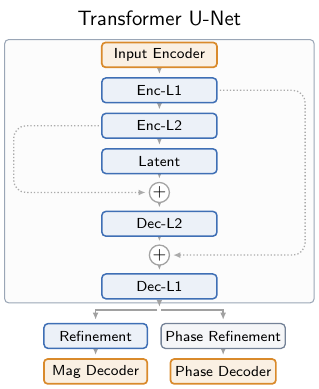}
    \caption{MUSE magnitude-pathway architecture probed in this work. A convolutional front end feeds six hierarchical transformer blocks across four stages---encoder, latent, decoder, and refinement---with four transformer layers per block, yielding 24 probed layers in total. Skip connections link encoder and decoder blocks at matched resolutions.}
    \label{fig:MUSE_architecture}
\end{figure}

MUSE \cite{lin24h_interspeech} is a transformer-convolutional SE model that follows a U-Net paradigm \cite{unet}, was trained on VoiceBank-DEMAND \cite{valentini2016voicebank}, and is organized into six hierarchical transformer blocks (Fig.~\ref{fig:MUSE_architecture}); activations are extracted from the first LayerNorm of each of its layers, yielding 24 probed layers in the magnitude pathway, the focus of this analysis. Throughout, we label the six blocks Enc-L1, Enc-L2, Latent, Dec-L2, Dec-L1, and Refinement, where the L-index denotes resolution level, so that the decoder traverses Dec-L2 before Dec-L1, and index the four layers of a block from $0$ to $3$, writing, for example, Dec-L1.3 for the last layer of the final decoder block.

At each probed layer~$\ell$, the per-utterance activation tensor $\mathbf{A}^{(\ell)} \in \mathbb{R}^{C \times T \times F}$, where $C$, $T$, and $F$ are the number of channels, time frames, and frequency bins, respectively, is reduced to a representation matrix $\mathbf{H}^{(\ell)} \in \mathbb{R}^{C \times F}$ by averaging over the time axis:
\begin{equation}
  \mathbf{H}^{(\ell)} = \frac{1}{T}\sum_{t=1}^{T} \mathbf{A}^{(\ell)}_{t}.
  \label{eq:pooling}
\end{equation}
Activations are extracted at each of the 24 transformer layers, yielding a single $C \times F$ matrix per utterance per layer. This time-averaging yields an utterance-length-invariant representation.

\textit{Additional probed architectures.} Two further models are probed under the same procedure, and are described with their results in Section~\ref{generalization_across_architecture}. For MP-SENet~\cite{mpsenet} we probe 8 layers: the first LayerNorm of each time- and frequency-transformer sub-block; its representation matrix is $[C, F]$, as in MUSE. For Demucs~\cite{demucs} we probe 11 layers: the output of each encoder block, decoder block, and the recurrent bottleneck; its representation matrix is $[T, C]$, time frames $\times$ channels. The sample axis of the similarity measure below is channels for the two spectral-domain models and time frames for Demucs.

\textit{Dereverberation fine-tuning.} Because the pretrained MUSE checkpoint was trained on non-reverberant speech, we fine-tuned it on convolutive mixtures to obtain the reverb-competent model probed under reverberation in Section~\ref{sec:experimental_setup}. Fine-tuning used 11{,}572 training utterances convolved on-the-fly with approximately 758 RIR-Mega training RIRs after reserving the validation set, with all parameters unfrozen and the original training loss retained. Optimization used AdamW ($\beta_1=0.8$, $\beta_2=0.99$) with a learning rate of $10^{-4}$, per-epoch exponential decay ($\gamma=0.99$), and a batch size of 28, on a 50-epoch schedule.

The checkpoint saved at epoch~48 attained the highest held-out PESQ improvement among the evaluated checkpoints and is the one probed; its dereverberation scores are reported alongside those of the other two architectures in Section~\ref{sec:reverb_ft}. The released code includes the full training configuration.

MP-SENet and Demucs were fine-tuned for dereverberation on the same convolutive mixtures, with all parameters unfrozen and each model's original training loss retained. MP-SENet used AdamW with the same settings as above, a learning rate of $10^{-4}$, an effective batch size of 28 via gradient accumulation, and a 50-epoch schedule; we probe the epoch-50 checkpoint. Demucs used the same optimizer, decay and batch size at a learning rate of $3\times10^{-4}$ on a 60-epoch schedule, together with a second run at $10^{-4}$ on a 50-epoch schedule matching the MP-SENet settings.

\subsection{Experimental Setup}
\label{sec:experimental_setup}

Clean utterances are drawn from the VoiceBank corpus, specifically the 824 utterances that constitute the VoiceBank-DEMAND test set (16\,kHz)~\cite{valentini2016voicebank, thiemann2013demand}, held out from the training data of the probed checkpoints. Rather than relying on the pre-mixed test set, we regenerate all mixtures by pairing each utterance with each DEMAND noise recording at the target integer SNRs, enabling evaluation under controlled degradation levels. The per-layer CKA at each of the 41 SNR levels is averaged over all $824 \times 18 = 14{,}832$ mixtures.

The VoiceBank-DEMAND test split includes only two speakers, so to apply speaker-level inference from Section~\ref{sec:perceptual}, we increase speaker diversity by adding two other speakers from the VoiceBank-DEMAND training set. Together, the two sweeps provide the four speakers analyzed in Section~\ref{sec:perceptual}; every other quantity reported in this paper is obtained from the sweep based on the VoiceBank-DEMAND test set described above.

For the reverberation experiments, since probing for dereverberation requires models that were trained for the task, we fine-tuned all three models for dereverberation, using the RIR-Mega dataset~\cite{rirmega} and clean utterances drawn from the training set of VoiceBank-DEMAND; the fine-tuning schedules for all three architectures are given in Section~\ref{sec:probed_model}.
These fine-tuned checkpoints are then probed with reverberant speech, based on the clean utterances of the VoiceBank-DEMAND test set, paired with RIRs from the AIR (Aachen Impulse Response) dataset~\cite{airdataset, airdataset_phone}, spanning six room types and a total of 88 RIRs. To sweep $C_{50}$ (Section~\ref{sec:degradation_model}) in a controlled manner, each RIR $h$ is partitioned into an early component $h_e$ (first 50\,ms) and a late tail $h_l$ (beyond 50\,ms). The late tail is then rescaled by a gain factor $g$ such that the modified RIR 
\begin{equation}
  h' = h_e + g \cdot h_l
  \label{eq:c50_scaling}
\end{equation}
targets the desired $C_{50}$. This procedure preserves the original RIR's early-reflection structure while adjusting only the late-reverberation energy. Thirteen target $C_{50}$ levels, linearly spaced from $-5$ to $25$\,dB, are evaluated per RIR, and the regression analyses in Section~\ref{generalization_to_reverb} are indexed by these nominal target levels. For each utterance, five RIRs are sampled at random from the pool and convolved at each target $C_{50}$. Across the full probing set, all 88 AIR RIRs are represented. Reference activations are computed using the same RIR with $g$ set such that $C_{50} = 50$\,dB, thereby preserving early reflections, which have been shown to benefit speech perception~\cite{10.1121/1.4789895}, while effectively removing late reverberation from the reference condition.

\subsection{Analysis Tools}

\subsubsection{Centered Kernel Alignment}
\label{sec:cka}

CKA is a scalar measure of representational similarity between two matrices of activations that share one dimension. We use the linear variant~\cite{kornblith2019similarity}, which is invariant to orthogonal transformations of the feature basis and to isotropic scaling.
Representational-similarity measures of this family, including linear CKA and CCA-based variants such as SVCCA and PWCCA, are widely used to analyze how representations evolve across the layers of deep speech and language models~\cite{pasad2021layerwise, chung2021similarity}. Formally, for two matrices $\mathbf{X} \in \mathbb{R}^{n \times p}$ and $\mathbf{Y} \in \mathbb{R}^{n \times q}$ with $n$ aligned samples, the linear gram matrices $\mathbf{K} = \mathbf{X}\mathbf{X}^\top$ and $\mathbf{L} = \mathbf{Y}\mathbf{Y}^\top$ are centered with $\mathbf{J}_n = \mathbf{I}_n - \tfrac{1}{n}\mathbf{1}\mathbf{1}^\top$, and linear CKA is defined as:
\begin{equation}
  \mathrm{CKA}(\mathbf{X}, \mathbf{Y})
  \;=\;
  \frac{\langle \mathbf{J}_n \mathbf{K} \mathbf{J}_n,\; \mathbf{J}_n \mathbf{L} \mathbf{J}_n \rangle_F}{\|\mathbf{J}_n \mathbf{K} \mathbf{J}_n\|_F \cdot \|\mathbf{J}_n \mathbf{L} \mathbf{J}_n\|_F}
  \;\in\; [0, 1].
  \label{eq:cka}
\end{equation}

Identical representations yield $\mathrm{CKA} = 1$ and uncorrelated ones yield $\mathrm{CKA} = 0$. In our work, at each probed layer $\ell$ we instantiate Eq.~\ref{eq:cka} with $\mathbf{X} = \mathbf{H}^{(\ell)}_x$, the representation matrix (Eq.~\ref{eq:pooling}) of an utterance under the clean reference $x$, and $\mathbf{Y} = \mathbf{H}^{(\ell)}_y$, the representation matrix of the same utterance under the degraded input $y$ at degradation level $s$. Both lie in $\mathbb{R}^{C \times F}$, so $n = C$ and channels play the role of aligned samples while frequency bins play the role of features; for Demucs, whose representation matrix is $[T, C]$, time frames play that role instead (Section~\ref{sec:probed_model}). The per-layer and degradation CKA score, denoted $\mathrm{CKA}(\ell, s)$, is then defined as the mean of $\mathrm{CKA}(\mathbf{H}^{(\ell)}_x, \mathbf{H}^{(\ell)}_y)$ over all mixtures at degradation level $s$.

The linear estimator of Eq.~\ref{eq:cka} is biased at finite sample size. To check that the depth profile is not an artifact of that bias, we recomputed it with the unbiased HSIC estimator of Song et~al.~\cite{song2012hsic}, in the form used for CKA-based network comparison by Nguyen, Raghu and Kornblith~\cite{nguyen2021wide}. This ablation was run on its own reduced grid, 150 utterances $\times$ 9 SNR levels $\times$ 2 DEMAND noises, with non-pooled activations re-extracted for the purpose; on that grid the unbiased estimator reproduces the biased one nearly exactly, at a per-layer slope correlation of $0.98$ and with an identical peak layer.

\subsubsection{Linearization of CKA Profiles}
\label{sec:linearization}

Each probed layer thus produces a $\mathrm{CKA}(\ell, s)$ curve over the swept degradation levels $s$. To summarize these curves compactly and compare them across layers and degradation types, we fit a first-order linear model per layer:
\begin{equation}
  \widehat{\mathrm{CKA}}(\ell, s) = \alpha_\ell + \beta_\ell \cdot s,
  \label{eq:regression}
\end{equation}
where $s$ denotes degradation level in dB (SNR or $C_{50}$). The slope $\beta_\ell$ quantifies how rapidly the representation at layer $\ell$ changes with degradation level and serves as a measure of \emph{sensitivity}. The intercept $\alpha_\ell$, which corresponds to $\mathrm{CKA}$ at $s = 0\,\text{dB}$, captures representational similarity under adverse conditions and serves as a measure of \emph{robustness}. The pair $(\alpha_\ell, \beta_\ell)$ thus provides a compact two-parameter profile for each layer, enabling direct comparison across degradation types and architectures.

Throughout, $R^2$ refers to the per-layer fit of Eq.~\ref{eq:regression} to the \emph{mean} degradation-level curve, the quantity plotted in every profile figure below.
Its adequacy varies across models, so we quote it where relevant rather than in aggregate.
Confidence intervals on $\alpha_\ell$ and $\beta_\ell$ are obtained by non-parametric bootstrap of the mean degradation-level curve, resampling hierarchically over the probing material (1{,}000 resamples, 2.5/97.5 percentiles)

\subsubsection{Quality-Association Analysis}
\label{sec:quality_method}

For each utterance and output metric $M$, we compute the improvement, namely:
\begin{align}
  \Delta M = M(\hat{x}, x) - M(y, x)
  \label{eq:delta_metric}
\end{align}
where, following the notation of Section~\ref{sec:degradation_model}, $x$ is the clean utterance and $y$ is the degraded input (Eq.~\ref{eq:additive_noise}), and $\hat{x}$ denotes the SE model's enhanced output, an estimate of $x$. We use PESQ~\cite{pesq}, STOI~\cite{taal2011stoi} and SI-SDR~\cite{leroux2019sisdr}. Since both CKA and $\Delta M$ are strongly driven by SNR, we control for this confound via within-group centering. For each layer $\ell$ and SNR level $s$, we subtract the group mean from both quantities before computing correlations. This corresponds to residualizing both variables with respect to the categorical SNR factor \cite{kutner2005applied}:
\begin{align}
  \mathrm{CKA}^c_{\ell_i} &= \mathrm{CKA}_{\ell_i} - \mathbb{E}[\mathrm{CKA} \mid \ell_i, s_i] \\ 
  \Delta M^c &= \Delta M - \mathbb{E}[\Delta M \mid s_i]
  \label{eq:centering}
\end{align}

We estimate the conditional expectations with sample means within each group, and compute per-layer Pearson correlations on the centered quantities. This approach removes any mean-level SNR effect without imposing a parametric form (e.g., linear dependence). CKA is centered per $(\ell, s)$, while $\Delta M$ is centered per $s$ only.

\subsubsection{Saturation Spread}
\label{sec:spread}

The two parameters of Eq.~\ref{eq:regression} are not independent summaries of a layer, and any reading of their relationship has to account for that. CKA is bounded above, and each layer's curve is anchored near its clean-reference value at the mild end of the sweep. Write $c^{\mathrm{high}}_\ell$ for the measured mean CKA at the mildest level of the sweep, and $s^{\mathrm{high}}$ for that level's coordinate: $30$\,dB for SNR and $25$\,dB for $C_{50}$. Since $\alpha_\ell$ is the value of the fit at $s = 0$\,dB, perfectly linear curves would satisfy $\beta_\ell = (c^{\mathrm{high}}_\ell - \alpha_\ell)/s^{\mathrm{high}}$ exactly. The across-layer correlation between intercept and slope would then be
\begin{equation}
  r(\alpha, \beta) = \frac{r_c f - 1}{\sqrt{f^2 - 2 r_c f + 1}},
  \qquad f = \frac{\mathrm{sd}(c^{\mathrm{high}})}{\mathrm{sd}(\alpha)},
  \label{eq:identity}
\end{equation}
with $r_c = r(\alpha, c^{\mathrm{high}})$, the across-layer Pearson correlation between the intercept and the clean-end value. The ratio $f$, which we call the \emph{saturation spread}, measures how widely the layers differ in where they saturate: it is the spread of clean-end values across layers, taken relative to the spread of the intercepts. Layers that all saturate to the same value give $f \approx 0$; layers whose clean ends differ give a larger $f$. Here $c^{\mathrm{high}}$ enters only as a diagnostic input to $f$ and is not itself reported as a summary of a layer. As $f \rightarrow 0$, the case in which every layer saturates to the same clean-end value, Eq.~\ref{eq:identity} forces $r(\alpha,\beta) \rightarrow -1$ whatever the data contain.

The consequence is not that the tradeoff is an artifact but that its evidential weight varies across measurements and can be read off. When $f$ is small, a strong negative correlation follows largely from Eq.~\ref{eq:identity}; when $f$ is large, the two parameters vary separately across layers, and a negative correlation is an empirical finding about the network. We therefore report $f$ alongside every correlation (Table~\ref{tab:tradeoff}), together with the variance share of the leading singular component of the layer-mean-centered $\text{layer} \times \text{level}$ CKA matrix, which measures the same thing from the data side: sweeps in which all layers saturate together are effectively one-dimensional across layers. Each row of Table~\ref{tab:tradeoff} is recomputed independently on that model's per-layer mean degradation-level curve, at 824-utterance scale.

\subsubsection{Diffusion Maps}
\label{sec:diffusion_maps_method}
We employ diffusion maps to complement CKA. Diffusion maps account for an entire set of representations at once, e.g., across all degradation values for a given block, and extract their joint intrinsic structure in a geometrically interpretable way.

For each layer~$\ell$ and degradation level~$s$, the centroid representation is computed by averaging over all $N_s$ utterances at that level:
\begin{equation}
  \bar{\mathbf{H}}^{(\ell)}_s = \frac{1}{N_s}\sum_{i=1}^{N_s} \mathbf{H}^{(\ell)}_{s,i} \;\in\; \mathbb{R}^{C \times F}.
  \label{eq:centroid}
\end{equation}
where $\mathbf{H}^{(\ell)}_{s,i}$ is the representation matrix (Eq.~\ref{eq:pooling}) of the $i$-th utterance under degradation level~$s$. Each centroid is vectorized to $\mathbb{R}^{C \cdot F}$, yielding a set of $S$ points (one per degradation level).

Given $S$ points in a fixed high-dimensional space, diffusion maps constructs a Gaussian affinity matrix $\mathbf{W} \in \mathbb{R}^{S \times S}$ with adaptive bandwidth, normalizes it into a row-stochastic transition matrix $\mathbf{P}$, and embeds the points via the leading eigenvectors and eigenvalues of $\mathbf{P}$, raised to the diffusion time $t$, into a low-dimensional, embedded space~\cite{coifman2006diffusion}. We use $t = 0.5$ for the per-layer views and $t = 5$ for the across-layer view.
The key property of diffusion maps is that for every two points, their Euclidean distance in the embedded space approximates their diffusion distance in the original space. The diffusion distance between two points integrates over all paths connecting them, weighted by transition probabilities between points in the set, rather than measuring direct pairwise proximity alone. Diffusion distance thereby measures proximity according to the local intrinsic structure of the high-dimensional point cloud~\cite{nadler2006diffusion}.

We apply diffusion maps twice independently. First, we fix a probed block and embed its centroids across all SNR levels, producing pairwise diffusion distances between degradation conditions within that block. In the second step, we fix a degradation level and embed the centroids across all probed layers, producing pairwise distances between layers at a given SNR.

\section{Results}
\label{sec:results}

We present the analysis in five stages. First, we establish the core depth-organized profile under additive noise using CKA and linear regression. Second, we relate it to output-level quality. Third, we test whether it generalizes across degradation types and across architectures, reporting the saturation spread alongside each correlation. Fourth, we show that it is induced by training rather than by architecture, and that adapting a trained model to a new degradation type preserves it. Finally, we view the same representations geometrically using diffusion maps.

\subsection{The Robustness-Sensitivity Tradeoff Under Noise}
\label{sec:noise_tradeoff}
\begin{figure}[t]
  \centering
  \includegraphics[width=\columnwidth]{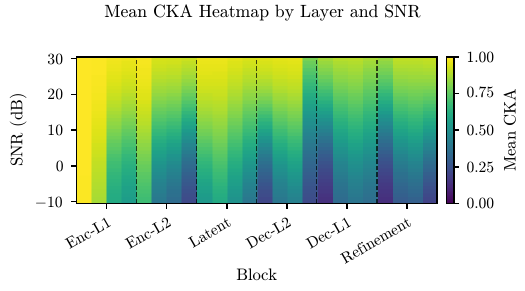}
  \caption{Layer-wise CKA between clean and noisy MUSE activations across the SNR sweep. Horizontal axis: the 24 probed layers, grouped by block (encoder, latent, decoder, refinement); vertical axis: input SNR from $-10$ to $30$\,dB; cell color: CKA averaged over utterances and noise types. Early encoder layers form a near-uniform high-similarity band across all SNRs, while later layers develop a clear gradient along the SNR axis.}
  \label{fig:cka_heatmap}
\end{figure}

We next examined how the similarity between clean and noisy MUSE activations varies with network depth and input SNR, over the full sweep of Section~\ref{sec:experimental_setup}. Figure~\ref{fig:cka_heatmap} reports layer-wise CKA across the SNR sweep, averaged over utterances and noise types. The resulting pattern is highly depth-dependent.

The first encoder layers maintain high similarity to their clean-reference activations across the full SNR range, appearing as a near-uniform high-CKA band. In deeper layers, CKA varies more strongly with SNR, producing a clear gradient along the SNR axis. The latent block shows an intermediate pattern, whereas the decoder and refinement layers exhibit the largest SNR-dependent changes. These results show that sensitivity to the degradation condition increases with depth, while early encoder representations remain comparatively stable under noise.

\begin{figure}[t]
  \centering
  \includegraphics[width=\columnwidth]{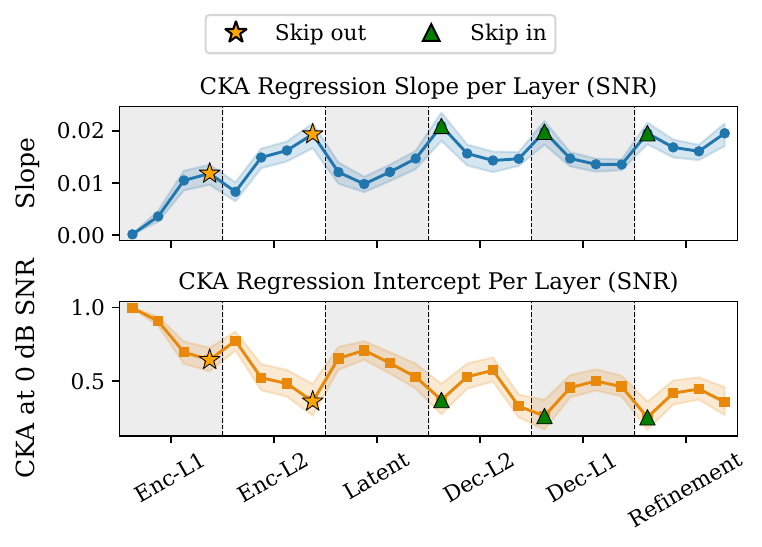}
  \caption{Per-layer robustness-sensitivity profile of MUSE under additive noise. Top: slope $\beta_\ell$ (sensitivity); bottom: intercept $\alpha_\ell$ (robustness, CKA at $0$\,dB), both from the per-layer linear fit of CKA vs.\ SNR (Eq.~\ref{eq:regression}). Shaded bands are $95\%$ bootstrap CIs, resampling hierarchically over noise types and utterances (1{,}000 replicates); alternating bands mark the six blocks.
 }
  \label{fig:regression}
\end{figure}

Linearization of the CKA-SNR relationship via Eq.~\ref{eq:regression} for each layer reveals finer patterns within each block (Fig.~\ref{fig:regression}). First, sensitivity increases monotonically with depth within each encoder block but resets at block boundaries.
In the decoder and refinement blocks, the within-block trend reverses: the first layer of each block exhibits the highest sensitivity. 
Sensitivity then decreases with depth through the decoder blocks, but not through the refinement block, whose first and last layers exhibit equal sensitivity. We therefore restricted the decreasing claim to the decoder blocks. Second, the intercept mirrors this pattern inversely, falling from $\alpha_\ell = 0.99$ at the first encoder layer to $0.25$ at the first refinement layer. Third, layers at decoder skip-connection boundaries produce local slope maxima and exhibit the highest sensitivity values in the network. The fits are strong throughout for MUSE under additive noise, $R^2 \geq 0.96$.

The profile is not driven by the choice of probing noise. The eighteen DEMAND recordings are divided by their role in VoiceBank-DEMAND into the five that constitute its test split, the out-of-distribution set for the pretrained MUSE, the eight that constitute its training split, and the five that enter neither. Fitting the profile independently on the first two of these subsets, over the same 824 utterances and the same 41 SNR levels, gives per-layer slopes that agree at $r = 0.9812$ and $\rho = 0.9757$ across the 24 probed layers, both placing the maximum at the same layer, Dec-L2.0. Mean $\beta_\ell$ is $0.0145$ on the held-out five against $0.0136$ on the training eight, but this difference does not track familiarity: the five environments that enter neither split, and are therefore as unseen by the pretrained model as the held-out five, sit at $0.0135$.

\subsection{Connection to Perceptual Quality}
\label{sec:perceptual}
\begin{figure}[t]
  \centering
  \includegraphics[width=\columnwidth]{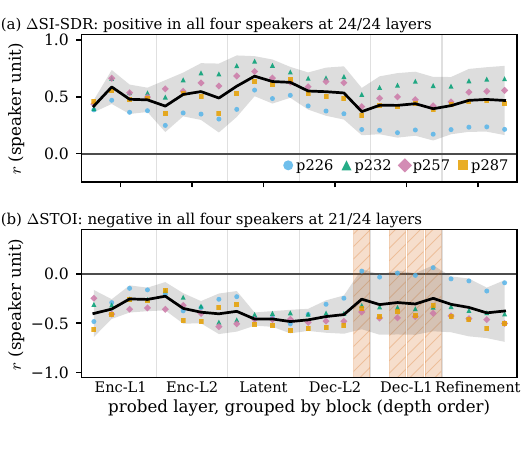}
  \caption{The quality association in MUSE with the speaker as the sampling unit, over four speakers. Black: the speaker mean; grey band: its $95\%$ $t$-interval, describing spread rather than testing; coloured dots: speakers; red columns: layers whose band covers zero. (a) For $\Delta$SI-SDR, all four are positive at every layer. (b) For $\Delta$STOI all four are negative except at Dec-L2.3, Dec-L1.1 and Dec-L1.3.}
  \label{fig:speaker_level}
\end{figure}

We examine whether these internal representational trends carry information on perceptual quality, using the improvement $\Delta M$ and the within-SNR centering of Section~\ref{sec:quality_method}.
We take the speaker as the sampling unit. At the deepest first-decoder layer, an utterance-cluster bootstrap gives an interval of width $0.041$, while resampling speakers gives one of $0.505$ at the same layer, twelve times wider and spanning zero. Four speakers cannot carry population-level inference, so we report per-speaker values, their sign agreement, and their spread (Fig.~\ref{fig:speaker_level}).

$\Delta$\textbf{SI-SDR} shows the most consistent association: it is positive in all four speakers at all 24 probed layers, spanning $0.17$ to $0.60$ at Dec-L1.3 ($r = 0.40$, $[0.12, 0.68]$). $\Delta$\textbf{STOI} is negative in all four speakers at most of the layers, over the encoder, the latent block, and the early decoder, but not at Dec-L2.3, Dec-L1.1, or Dec-L1.3, where a single speaker reverses the sign; the band covers zero there and at Dec-L1.2. We therefore do not describe it as consistently negative across depth.

\begin{figure}[t]
  \centering
  \includegraphics[width=\columnwidth]{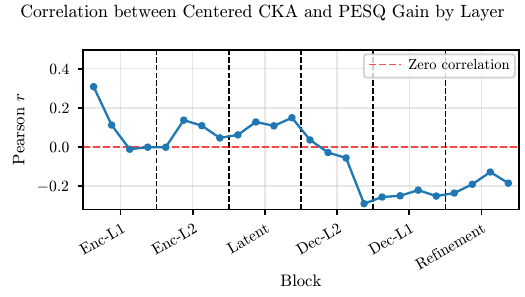}
  \caption{Per-layer Pearson correlation between CKA and PESQ improvement $\Delta\mathrm{PESQ}$ after within-group centering on SNR (Eq.~\ref{eq:centering}), computed over utterances. Correlations are mildly positive across the encoder and latent blocks, largest at the first encoder layer, turn increasingly negative through Dec-L2, and saturate through Dec-L1 and refinement. The per-layer values are exact, but the utterance is not the correct sampling unit for them: with the speaker as the unit, only the early positive association is shared by all four speakers; in the decoder lobe their signs split (Section~\ref{sec:perceptual}).}
  \label{fig:pesq_corr}
\end{figure}

For $\Delta$\textbf{PESQ}, the choice of sampling unit changes the conclusion, and Fig.~\ref{fig:pesq_corr} should be read accordingly. At the utterance level, the figure shows small positive correlations across the encoder and latent blocks, largest at the first encoder layer, and an increasingly negative lobe through the first decoder block. The underlying per-layer values are exact, but the reading drawn from them is not: with the speaker as the sampling unit the four speakers do not agree in sign anywhere in that lobe (deepest Dec-L2 layer: signs split three to one, mean $r = -0.21$, spread $[-0.46, +0.05]$), whereas in the early encoder and latent blocks all four are positive (first encoder layer: mean $r = +0.29$, spread $[+0.20, +0.38]$). We draw no architectural conclusion from the decoder-side sign, which two speakers cannot resolve.

\subsection{Generalization Across Degradation Types}
\label{generalization_to_reverb}

\begin{figure}[t]
  \centering
  \includegraphics[width=\columnwidth]{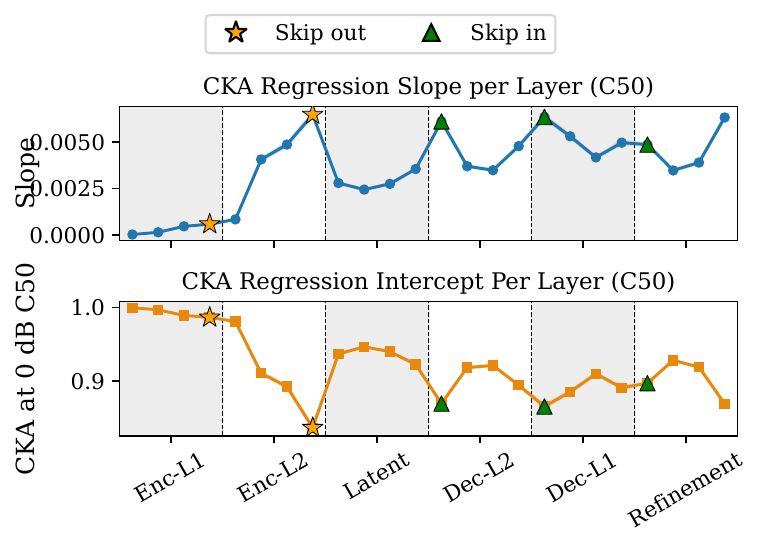}
  \caption{Per-layer robustness-sensitivity profile of MUSE under reverberation, on 824 utterances $\times$ 13 nominal target $C_{50}$ levels ($-5$ to $25$\,dB) over the AIR pool of 88 RIRs spanning six rooms. Top: slope $\beta_\ell$ (sensitivity); bottom: intercept $\alpha_\ell$ (robustness, CKA at $0$\,dB), both from the per-layer linear fit of Eq.~\ref{eq:regression}. Shaded bands are $95\%$ bootstrap CIs (Section~\ref{sec:linearization}); alternating bands mark the six blocks. The layout matches Fig.~\ref{fig:regression}.}
  \label{fig:c50_regression}
\end{figure}

To determine whether the tradeoff established in Section~\ref{sec:noise_tradeoff} is specific to additive noise or reflects a more general response to input degradation, we repeated the regression analysis under controlled reverberation, sweeping $C_{50}$ over a pool of 88 RIRs spanning six rooms, with each architecture probed at its dereverberation fine-tune (Section~\ref{sec:experimental_setup}). 
Figure~\ref{fig:c50_regression} reports the resulting MUSE profile, in the layout of the noise-axis figure; the corresponding per-layer statistics for MP-SENet and Demucs are shown in Fig.~\ref{fig:scatter} and Table~\ref{tab:tradeoff}. 
As in the noise setting, early encoder layers remain close to their reference representations across the full $C_{50}$ range. Deeper layers, particularly in the decoder and refinement blocks, exhibit increasing sensitivity to degradation. The same depth-dependent progression of slopes (sensitivity) emerges, with intercepts following the inverse pattern. The largest slopes again concentrate at skip-connection boundaries, though under reverberation the network maximum falls at the encoder skip source rather than at a decoder junction.

Sensitivity increases with depth in both spectral-domain models on both axes, most strongly for MP-SENet under noise, $\rho(\text{depth}, \beta) = +0.833$, where the deepest frequency-transformer sublayer is the most sensitive layer of the network; the per-layer fits are tabulated in the released analysis artifacts. Demucs, discussed in Section~\ref{generalization_across_architecture}, distributes its sensitivity differently but is equally stage-structured.

What a model does under noise largely anticipates what it does under reverberation. Correlating each architecture's noise-axis profile with its reverberation-axis profile across that architecture's own layers, MUSE agrees on both parameters, a rank correlation of $+0.837$ $[+0.582, +0.942]$ for sensitivity and $+0.864$ $[+0.664, +0.957]$ for robustness, with no single layer carrying the result, and MP-SENet preserves its robustness ordering, $+0.833$ $[+0.300, +1.000]$. Demucs is organized around its bottleneck rather than depth, and the bottleneck carries across: the LSTM is the most robust layer under both degradations, first of eleven on each axis, with intervals disjoint from the runner-up.

\subsection{Generalization Across Architectures}
\label{generalization_across_architecture}

To determine whether the tradeoff reflects a property of the enhancement objective or a particular architectural choice, we applied the same probing procedure to two additional architectures that differ substantially from MUSE in design:

\begin{itemize}
  \item \textbf{MP-SENet}~\cite{mpsenet} is a frequency-domain model that processes magnitude and phase spectra in parallel via alternating time- and frequency-transformer blocks. We use the authors' released DNS checkpoint.
  \item \textbf{Demucs}~\cite{demucs} is a time-domain convolutional-recurrent model with a symmetric encoder-decoder structure connected by a unidirectional LSTM bottleneck. We use the DNS64 checkpoint. Its 11 probed layers and $[T, C]$ representation matrix are those of Section~\ref{sec:probed_model}.
\end{itemize}

The probed layer count differs across models since we probe at structural boundaries, and the architectures expose different numbers of them: normalization points within transformer blocks for MUSE and MP-SENet, and block outputs together with the recurrent bottleneck for Demucs, which has no comparable internal boundary. The exact module paths are listed in the released artifacts. We do not compute CKA between architectures; instead, we probe each model independently and compare the resulting $(\alpha_\ell, \beta_\ell)$ profiles.

\begin{figure}[t]
  \centering
  \includegraphics[width=\columnwidth]{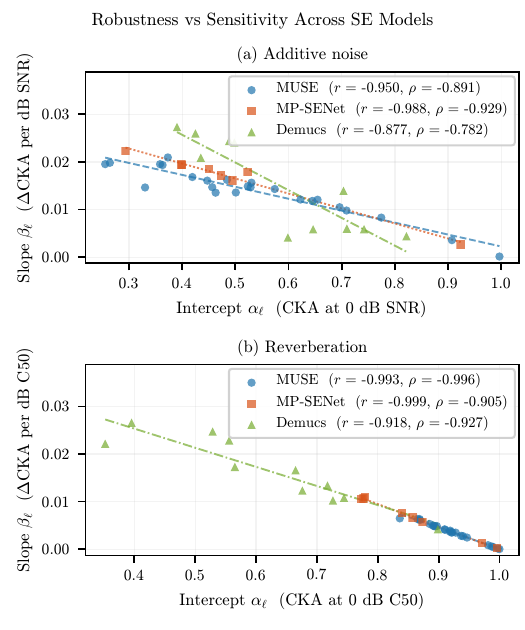}
  \caption{Robustness-sensitivity tradeoff across three SE architectures and both degradation axes. Each point represents one probed layer, plotted by its CKA-degradation intercept $\alpha_\ell$ (robustness, horizontal axis) against its slope $\beta_\ell$ (sensitivity, vertical axis), with fits independently obtained for MUSE, MP-SENet, and Demucs; dashed lines are per-model least-squares fits. The correlation is negative in all six cases; Table~\ref{tab:tradeoff} reports each with the saturation spread that scopes it.}
  \label{fig:scatter}
\end{figure}

Both architectures are as strongly non-uniform across depth as MUSE, but they organize that non-uniformity differently from it and from each other. Figure~\ref{fig:scatter} summarizes those profiles, and Table~\ref{tab:tradeoff} reports the across-layer correlation between the per-layer intercepts and slopes for each of the six cases, over that model's $n_\ell$ probed layers, by Pearson and by Spearman~\cite{spearman1904} alike, each alongside the saturation spread $f$ and the leading-component share PC1 of Section~\ref{sec:spread} that scope it.

\begin{table}[t]
\centering
\footnotesize
\setlength{\tabcolsep}{3pt}
\caption{Robustness-Sensitivity Tradeoff and its Scope, per Architecture and Degradation Axis. Symbols are defined in Section~\ref{sec:spread}. SNR Rows are the Pretrained Checkpoints and $C_{50}$ Rows are the Dereverberation Fine-Tunes (Section~\ref{sec:experimental_setup}).}
\label{tab:tradeoff}
\begin{tabular}{@{}llccccc@{}}
\toprule
Model & Axis & $n_\ell$ & $r(\alpha,\beta)$ & $\rho(\alpha,\beta)$ & $f$ & PC1 (\%) \\
\midrule
MUSE     & SNR & 24 & $-0.950$ & $-0.891$ & $0.275$ & $97.54$ \\
MUSE     & $C_{50}$ & 24 & $-0.993$ & $-0.996$ & $0.037$ & $96.52$ \\
MP-SENet & SNR &  8 & $-0.988$ & $-0.929$ & $0.137$ & $98.48$ \\
MP-SENet & $C_{50}$ &  8 & $-0.999$ & $-0.905$ & $0.025$ & $99.23$ \\
Demucs   & SNR & 11 & $-0.877$ & $-0.782$ & $0.956$ & $76.32$ \\
Demucs   & $C_{50}$ & 11 & $-0.918$ & $-0.927$ & $0.167$ & $83.58$ \\
\bottomrule
\end{tabular}
\end{table}

By Eq.~\ref{eq:identity} the sign of $r(\alpha,\beta)$ turns on the product $r_c f$ alone, and is negative whenever $r_c f < 1$. Since $r_c = r(\alpha, c^{\mathrm{high}})$ is a correlation and cannot exceed $1$, a non-negative $r(\alpha,\beta)$ would require $f > 1$, which no row reaches: across layers $r_c$ is $+0.823$, $+0.853$ and $-0.327$ for MUSE, MP-SENet and Demucs under noise and $+0.860$, $+0.918$ and $+0.584$ under reverberation, so $r_c f$ nowhere exceeds $0.227$. A negative sign in all six cases is therefore not independent evidence, and we do not present it as one; it is a constrained descriptive summary of the per-layer profiles, and the profiles carry the result. Magnitude is less constrained. At a given $f$ the identity permits $|r(\alpha,\beta)|$ as small as $\sqrt{1-f^2}$, reached at $r_c = f$. Demucs under noise is the loosest row, at $f = 0.956$ and the lowest leading-component share of any row; its clean end varies across depth, so $\alpha_\ell$ and $\beta_\ell$ are separable there, and against a permitted floor of $0.294$ the measurement is $0.877$. That floor is exact only for a perfectly linear CKA curve, whose fitted line passes through the clean-end value; the five more saturated rows bend enough to fall below it, by $0.001$ to $0.068$.

\textit{The recurrent bottleneck is Demucs's most robust probed layer.} The three architectures are all strongly non-uniform across depth. Still, they distribute sensitivity differently: MUSE and MP-SENet grow more sensitive with depth, whereas Demucs grows less sensitive, a monotone decrease ($\rho(\text{depth}, \beta) = -0.727$ on the noise axis) that takes the form of an hourglass rather than a gradual falloff, with the encoder stack and the deep decoder both strongly condition-dependent and a single highly robust point between them. That point is the LSTM bottleneck, and at 824-utterance scale it is the network maximum of $\alpha_\ell$ on both degradation axes. Under reverberation it reaches $\alpha_\ell = 0.8986$ against $0.7444$ for the runner-up encoder block, a gap of $+0.154$ $[+0.151, +0.158]$; it is the arg-max of $\alpha_\ell$ in $1{,}000$ of $1{,}000$ utterance bootstrap resamples, and the arg-min of $\beta_\ell$ in $1{,}000$ of $1{,}000$. It replicates on the noise axis, reaching $\alpha_\ell = 0.821$ versus $0.741$ for the runner-up decoder block. In the tradeoff coordinates of Fig.~\ref{fig:scatter}a, it is the extreme robustness-dominant point of the Demucs cloud, in sharp contrast with the encoder blocks.

Demucs is also the model on which the linear summary is weakest, with a minimum $R^2$ of $0.441$ at {\small\texttt{decoder.3}} under noise. That minimum reflects the shape of the curve rather than a failure of the summary: {\small\texttt{decoder.3}}'s CKA-SNR curve spans only $0.094$ CKA over the whole $40$\,dB sweep, so a low $R^2$ reflects a nearly flat curve fitted by a nearly flat line. Dropping the worst-fitting layers would not be defensible, because an $R^2$ cut would preferentially delete the flat, robust layers and inflate the very depth trend the analysis measures, and no layer is accordingly excluded.

\subsection{Origin of the Profile in Training}
\label{sec:random_init}

A remaining possibility is that the profile is a property of the MUSE architecture itself, of its U-Net skip topology and time-pooling, rather than an effect of the enhancement objective. To separate the two, we apply the identical probing pipeline to an untrained MUSE, instantiated from the same configuration but with randomly initialized weights and no checkpoint loaded, over three random seeds.
The control runs on the noise axis, over the same sweep as the main analysis.

\begin{figure}[t]
  \centering
  \includegraphics[width=\columnwidth]{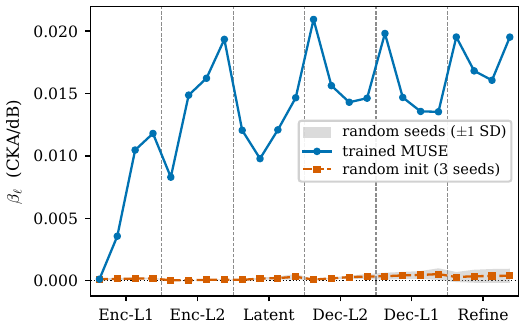}
  \caption{Random-initialization control. The identical probing pipeline applied to an untrained MUSE over three random seeds produces no comparable profile: CKA stays near unity at every layer and level, and the per-layer slopes are one to two orders of magnitude below the trained model's.}
  \label{fig:random_init}
\end{figure}

Without training, the profile does not appear (Fig.~\ref{fig:random_init}). Averaged over utterances, noise types, and seeds, CKA stays near unity at every layer and degradation level, with a minimum of $0.9802$. The per-layer sensitivity slopes are correspondingly small: per-seed maxima of $0.00028$, $0.00105$ and $0.00045$ against the trained model's peak $\beta_\ell$ of $0.02261$, one to two orders of magnitude larger; even the largest untrained value is a factor of $21.5$ lower. Random weights are not free of depth structure, and once both sides are ordered by depth that structure runs in the same direction as the trained model's: two of the three seeds give a positive rank correlation between depth and slope and the third none ($\rho(\text{depth}, \beta) = +0.43$, $+0.86$ and $-0.07$; the first two reach the $5\%$ level, the third does not). The trained model is distinguished by effect size rather than sign, along with skip-junction peaks that have no counterpart in the untrained network. Training therefore induces this organization, and it is not an artifact of the architecture or skip-connection concatenation.

The reverb-competent checkpoint of Section~\ref{sec:experimental_setup} is the endpoint of a 50-epoch run. The artifacts include CKA sweeps at ten intermediate checkpoints, at epochs $1$, $3$, $5$, $7$, $9$, $10$, $12$, $24$, $36$ and $48$, each over the 824-utterance sweep of Section~\ref{sec:experimental_setup} at the same 41 SNR levels and five DEMAND environments, so refitting Eq.~\ref{eq:regression} per layer at each lets us watch the profile form. On the noise axis, the characteristic organization is present at the first probed checkpoint and only sharpens: the $\beta_\ell$ profile's correlation with the final epoch-48 profile rises from $r = 0.76$ at epoch~1 to $r \geq 0.99$ by epoch~36. The pretrained checkpoint already correlates with the final profile at $r = 0.76$ on $\beta_\ell$ and $r = 0.73$ on $\alpha_\ell$. The change fine-tuning induces is concentrated in exactly the stages the profile labels as sensitive: between the pretrained and epoch-48 checkpoints, the mean absolute change in $\beta_\ell$ is $3\times$ larger, and in $\alpha_\ell$ $5.8\times$ larger, in the decoder and refinement blocks than in the first encoder block. We report this analysis only on the noise axis.

\subsection{Effect of Dereverberation Fine-Tuning}
\label{sec:reverb_ft}

The emergence analysis tracks one model along one fine-tuning run. To determine whether adapting a trained model to a new degradation type reorganizes where it is condition-dependent, we fine-tuned MP-SENet and Demucs for dereverberation on RIR-Mega~\cite{rirmega} convolutive mixtures under the schedules of Section~\ref{sec:experimental_setup}, and re-probed them together with the MUSE dereverberation fine-tuning described there. For Demucs we report two checkpoints: one at epoch~57 with learning rate $3\times10^{-4}$, and one at epoch~45 with learning rate $10^{-4}$, matched to the MP-SENet schedule. MP-SENet and Demucs both learn the task: on a held-out set of $4{,}120$ mixtures---824 utterances $\times$ five RIR draws from a pool of 202 held-out RIR-Mega impulse responses---dereverberation fine-tuning raises PESQ from $1.658$ on the reverberant input to $3.361$ for MP-SENet and $2.441$ for Demucs, the learning-rate-matched run reaching $2.339$, with STOI and SI-SDR improving in step: MP-SENet gains $19.6$\,dB of SI-SDR and Demucs $13.9$\,dB. The corresponding pretrained checkpoints reach PESQ $1.69$ and $1.73$ on the same set, gain under $1$\,dB of SI-SDR over the reverberant input, and fall below it on STOI, indicating that dereverberation competence is acquired during fine-tuning rather than inherited from the denoising checkpoints. The MUSE fine-tune of Section~\ref{sec:experimental_setup} was evaluated on a different held-out set of the same size, 824 utterances $\times$ five distinct RIR-Mega impulse responses of mean $C_{50}$ $12.0$\,dB, independent of the AIR probing RIRs, on which it reaches PESQ $=3.02$ and STOI $=0.944$~\cite{pesq, taal2011stoi} against PESQ $=2.17$ and STOI $=0.834$ for the denoising checkpoint. That set is less adverse than the 202-RIR split used for the other two architectures, so absolute values are not comparable across the two evaluations, and each architecture is read against its own pretrained arm.
MP-SENet is substantially stronger on every metric, while the two Demucs runs are close, so the difference between these two schedules does not explain the gap.
The impulse responses used here are held out from RIR-Mega, the corpus these models were fine-tuned on; the reverberation \emph{probing} of Section~\ref{generalization_to_reverb} instead uses the AIR six-room pool, a separate database serving a separate purpose, so the two are not larger and smaller versions of one collection.

For each architecture, we re-probed the pretrained and dereverberation fine-tuned checkpoints, using the epoch-57 run for Demucs under reverberation, over the same pool of 88 RIRs and the same 13 target levels, and the depth organization unchanged. The per-layer $\beta_\ell$ profile before and after fine-tuning correlates at $r = 0.91$ for MUSE, $0.81$ for MP-SENet and $0.98$ for Demucs, and the intercept profiles at $0.85$, $0.84$ and $0.90$; mean absolute changes are $0.0020$, $0.0029$ and $0.0015$ in $\beta_\ell$, against slopes reaching $0.02$. Adapting a denoising model to a new degradation type shifts the profile magnitude more than its shape.

\subsection{A Geometric View}
\label{diffusion}

CKA regression quantifies how each layer's representations differ from clean references as a scalar function of degradation level. To complement it, we apply diffusion maps (Section~\ref{sec:diffusion_maps_method}).
We carry out this analysis for MUSE along the noise axis. Each centroid of Eq.~\ref{eq:centroid} averages the 824 utterances of Section~\ref{sec:experimental_setup} and is then averaged over five held-out DEMAND environments, giving one point per SNR level and $S = 41$ points per block. The Gaussian affinity of Section~\ref{sec:diffusion_maps_method} takes the median pairwise squared distance as its bandwidth, and a $99\%$ cumulative eigenvalue-energy cutoff sets the embedding dimension.

\subsubsection{Diffusion Distances Across SNR}

\begin{figure}[t]
  \centering
  \includegraphics[width=\columnwidth]{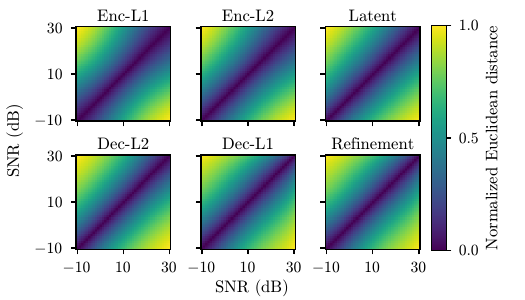}
  \caption{Pairwise diffusion distances between MUSE centroid representations across the SNR sweep of Section~\ref{diffusion}, one matrix per probed block. Row $i$, column $j$: diffusion distance between centroids at SNR $s_i$ and $s_j$; both axes ordered by SNR. Distances from the $30$\,dB reference grow monotonically as SNR decreases in every block. Encoder blocks show gradual gradients; decoder blocks transition more sharply and reach larger off-diagonal distances at low SNR.}
  \label{fig:diffusion}
\end{figure}

Figure~\ref{fig:diffusion} shows pairwise diffusion distances between the centroid representations across SNR levels for each probed block. Representations are ordered consistently by SNR: distances from the 30\,dB reference increase monotonically as SNR decreases, with absolute Spearman $|\rho|=1.00$ across all six probed blocks, so the ordering recovered by CKA is not specific to that scalar metric.

Magnitude is the more informative quantity here, and it is not implied by the ordering: a sweep can be perfectly ordered in every block while every block travels the same distance. Encoder blocks show gradual distance gradients, while decoder blocks transition more sharply and reach larger pairwise distances at low SNR. The same asymmetry appears as geometric displacement: within every block the centroids arrange along a monotonic arc in the first two diffusion coordinates, and the mean arc length is $3.03\times$ larger for decoder than for encoder blocks ($0.193$ against $0.064$), consistent with their larger sensitivity slopes $\beta_\ell$ on this model and axis.

\subsubsection{Diffusion Distances Across Layers}

\begin{figure}[t]
  \centering
  \includegraphics[width=\columnwidth]{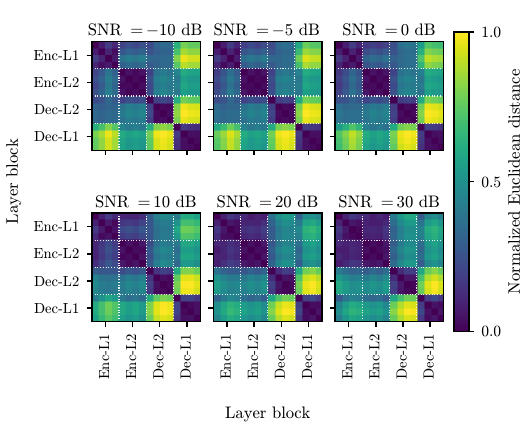}
  \caption{Pairwise Euclidean distances between diffusion-map centroids of 16 probed MUSE layers across the encoder and decoder blocks (Enc-L1, Enc-L2, Dec-L2, Dec-L1), one matrix per SNR. Layers are ordered by depth. At low SNR, the encoder and decoder layers form two tight clusters with the largest distance between them; as SNR increases, the inter-group gap shrinks monotonically. Layers with the largest slopes $\beta_\ell$ move the most across panels; those with the largest intercepts $\alpha_\ell$ stay put.}
  \label{fig:layer_distance_matrices}
\end{figure}

Switching axes, we fix the SNR and compute pairwise Euclidean distances between diffusion-map centroids across the probed encoder and decoder layers, producing one distance matrix per SNR level (Fig.~\ref{fig:layer_distance_matrices}). At low SNR, encoder layers cluster tightly among themselves, while decoder layers form a separate cluster, with the largest distances occurring between the two groups. As SNR increases, these inter-group distances shrink monotonically, and the matrices become nearly uniform, consistent with the network approaching the clean-aligned configuration. 
Layers with the largest sensitivity slopes $\beta_\ell$ are the same ones that move the most across the SNR panels, while layers with the largest robustness intercepts $\alpha_\ell$ remain in roughly the same neighborhood. Taken together, the scalar regression and the geometric analysis recover the same depth-dependent partitioning. Both are computed from the same activations, so this is a consistency check across similarity measures rather than independent evidence; it indicates that the partitioning does not depend on the choice of CKA.

\section{Discussion}
\label{sec:discussion}

The robustness-sensitivity profile provides a compact way to interpret how speech enhancement models distribute their computation across depth. Read together, the two quantities turn layer-wise CKA curves into an architectural diagnostic: high robustness and low sensitivity indicate representation-preserving stages, whereas lower robustness and higher sensitivity indicate stages that perform condition-dependent transformation.

Using this profile, MUSE separates into functional regions: the encoder is robustness-dominant, the latent block is intermediate, and the decoder and refinement blocks carry most of the degradation-dependent variation. The strongest local sensitivity appears near decoder skip-connection boundaries, suggesting that reintegrating encoder information is a key site where clean-aligned structure is transformed into an enhancement-specific representation.

This division of labor connects naturally to classical spectral enhancement, where the transformation applied to the input depends on local degradation. Under the standard Gaussian-additive-noise model, the Wiener suppression gain $G(\xi) = \xi/(1+\xi)$, where $\xi$ is the \textit{a priori} SNR at a time-frequency bin~\cite{scalart1996speech, malah_se_wiener}, serves as a baseline for spectral enhancement. As $\xi \to \infty$ (clean input), the gain saturates at $1$ and becomes locally insensitive to the input condition; as $\xi \to 0$, the gain approaches $0$ and changes more sharply with $\xi$. Optimal suppression is therefore inherently condition-dependent, with the strongest dependence under adverse SNR. The robustness-sensitivity profile suggests that the decoder and skip-connection junctions are the architectural regions where this condition-dependent processing is most concentrated, while successive decoder layers appear to refine the transformation with diminishing marginal sensitivity. Demucs offers a complementary reading: its recurrent bottleneck, the one place where information is carried across time through a fixed-width state, is also the one place least reshaped by the input condition, on both degradation axes.

The scope condition of Section~\ref{sec:spread} extends beyond the present models. Any probe that measures similarity to a clean reference while sweeping severity produces curves anchored near their clean-end values at the mild end, so any two-parameter summary of such curves will show a negative intercept-slope relationship whose strength depends on how uniformly the layers saturate. 
Reporting the saturation spread alongside the correlation requires a single additional number and turns the objection into a measurable quantity: it fixes both the sign the identity forces and the smallest $|r|$ it permits. The evidence for the tradeoff is therefore the depth-organized profile itself. This profile is invariant to any reordering of the layers and is not constrained by the saturation identity.

The CKA-quality analysis bounds this interpretation. After within-SNR centering, the residual correlations are modest and, more importantly, governed by a sampling unit that the standard test material barely supplies: two speakers per sweep. With the speaker as the unit, the four agree in sign on the SI-SDR and STOI associations, while they do not share the decoder-side PESQ lobe reported with apparent precision at utterance level. The robustness-sensitivity profile thus explains where the model changes with degradation severity, but it should not be read as a direct predictor of utterance-level quality improvement. This point extends beyond the present setting: severity-swept probing of quality associations on a two-speaker test set is pseudo-replicated, and utterance-level intervals for it overstate precision by roughly an order of magnitude.

Under reverberation, both inter-block and intra-block trends closely mirror those observed under additive noise in MUSE, and each architecture's own stage structure likewise carries over from noise to reverberation. The diffusion-map analysis yields the same partitioning from pairwise distances alone on MUSE and the noise axis, so depth organization is not a property of the CKA scalar. 
Transformer, convolutional-recurrent, and hybrid designs are all distinctly stage-structured; however, they position their sensitive stages differently. Section~\ref{generalization_across_architecture} elaborates on the accompanying intercept-slope sign, while depth-organized profiles carry the evidence.
Three results indicate that the organization is a product of training toward the enhancement objective rather than of architecture: a randomly initialized MUSE, probed identically, produces a near-flat profile, so the skip-junction maxima in particular are not artifacts of concatenation; intermediate checkpoints show the profile present at the first probed checkpoint and sharpening thereafter; and the change fine-tuning induces concentrates in exactly the stages the profile labels sensitive, the mean absolute change in $\beta_\ell$ being $3\times$ larger and in $\alpha_\ell$ $5.8\times$ larger in the decoder and refinement blocks than in the first encoder block. Adapting a trained model to a new degradation type does not reorganize the profile either: the dereverberation fine-tunes of Section~\ref{sec:reverb_ft} learn the task, and re-probing them under reverberation leaves the profile largely unchanged.

Several limitations qualify this interpretation. CKA measures representational similarity, not the exact computation performed by each layer, and profiles are comparable only within a fixed sample-axis convention, since that convention determines which invariance is measured. The present experiments cover controlled single-degradation axes rather than the full diversity of real acoustic environments, and three architectures are a small sample of the design space. These limitations bound the scope of the claim: the comparison to spectral enhancement provides one interpretation of why sensitivity may concentrate in later condition-dependent stages, while broader validation across training corpora, degradation mixtures, more speakers, and generative enhancement models remains an important next step.

\section{Conclusions}
\label{sec:conclusions}

We presented a probing study of speech enhancement models combining controlled degradation, CKA-based regression, and diffusion-based geometric analysis. Applied to MUSE, the robustness-sensitivity profile reveals a consistent depth-dependent organization: encoder layers remain comparatively stable across degradation conditions, whereas decoder and refinement layers carry most of the degradation-dependent variation, with pronounced sensitivity at the first layer of each decoder block, where encoder information is reintegrated through a skip connection. 

A negative robustness-sensitivity relationship persists under reverberation and appears in MP-SENet and Demucs on both degradation axes. However, the three architectures place their sensitive stages differently: Demucs inverts the depth trend and concentrates robustness in its recurrent bottleneck, the layer least reshaped by the input condition on either axis. What generalizes is the stage-structured organization itself, not its MUSE-specific layout. Because CKA saturates at the clean reference, that relationship is partly definitional, and we quantify how much with a saturation-spread statistic reported alongside every correlation; the tradeoff is most testable, and holds, in the one measurement whose layers do not saturate uniformly.

A randomly initialized model shows no such profile; the organization is present at the first fine-tuning checkpoint and sharpens thereafter, and the changes fine-tuning induces concentrate in exactly the stages the profile marks as sensitive, indicating that it is induced by the enhancement objective rather than by a particular topology. Fine-tuning a trained model for dereverberation leaves the profile largely in place (Section~\ref{sec:reverb_ft}).

The centered CKA-quality analysis must treat the speaker as the sampling unit; however, the standard test material provides only two speakers per sweep, and it is exploratory, and the decoder-side PESQ association does not survive the change of unit. The profile should therefore be interpreted as a condition-level diagnostic rather than a direct utterance-level predictor of quality. Future work should test whether this organization persists across larger training corpora, mixed acoustic degradations, and generative enhancement models, and whether a test set with enough speakers can resolve the quality association that the present one cannot. Together, these results complement output-level evaluation by exposing how enhancement models distribute representation-preserving and condition-dependent processing across depth.

\section*{Data Availability}
Code and reproducibility materials are available at \url{https://github.com/YairAmar/SE-Probe}, and model checkpoints are available at \url{https://huggingface.co/yairamr/SE-Probe-models}. The dataset repository \url{https://huggingface.co/datasets/yairamr/SE-Probe-data} carries the raw per-utterance CKA artifacts for the noise and reverberation sweeps, the per-layer fit tables, the diffusion-map artifacts, and the dereverberation fine-tuning artifacts.

\balance

\bibliographystyle{IEEEtran}
\bibliography{bibliography_v5}

\end{document}